\documentclass{emulateapj112604}
\usepackage{psfig}
\usepackage{apjfonts}

\journalinfo{The Astrophysical Journal, in press}
\slugcomment{Received 2005 May 22; accepted 2005 September 26}

\shortauthors{Camilo et al.}
\shorttitle{PSR~J1833--1034 in SNR~G21.5--0.9}

\begin{document}

%  Definitions

\def\ccpc{cm$^{-3}$\,pc}
\def\chandra{{\em Chandra\/}}
\def\ergs{ergs\,s$^{-1}$}
\def\HI{H\,{\sc i}}
\def\kms{km\,s$^{-1}$}
\def\mkpc{mJy\,kpc$^2$}
\def\msun{$M_\odot$}
\def\psr{PSR~J1833--1034}
\def\snr{G21.5--0.9}

\title{PSR~J1833--1034: DISCOVERY OF THE CENTRAL YOUNG PULSAR IN THE
SUPERNOVA REMNANT G21.5--0.9}

\author{F.~Camilo,\altaffilmark{1}
  S.~M.~Ransom,\altaffilmark{2}
  B.~M.~Gaensler,\altaffilmark{3,4}
  P.~O. Slane,\altaffilmark{3}
  D. R. Lorimer,\altaffilmark{5}
  J. Reynolds,\altaffilmark{6}
  R.~N.~Manchester,\altaffilmark{7}
  and S.~S.~Murray\altaffilmark{3} }
\altaffiltext{1}{Columbia Astrophysics Laboratory, Columbia University,
  550 West 120th Street, New York, NY~10027.}
\altaffiltext{2}{NRAO, 520 Edgemont Road, Charlottesville, VA~22903.}
\altaffiltext{3}{Harvard-Smithsonian Center for Astrophysics, 60 Garden
  Street, Cambridge, MA~02138.}
\altaffiltext{4}{Alfred P. Sloan Research Fellow.}
\altaffiltext{5}{University of Manchester, Jodrell Bank Observatory,
  Macclesfield, Cheshire, SK11~9DL, UK.}
\altaffiltext{6}{Australia Telescope National Facility, CSIRO,
  Parkes Observatory, P.O.~Box~276,  Parkes, NSW~2870, Australia.}
\altaffiltext{7}{Australia Telescope National Facility, CSIRO,
  P.O.~Box~76, Epping, NSW~1710, Australia.}

\begin{abstract}
We have discovered the pulsar associated with the supernova remnant \snr.
\psr, with spin period $P=61.8$\,ms and dispersion measure 169\,\ccpc,
is very faint, with pulse-averaged flux density of $\approx 70\,\mu$Jy
at a frequency of 1.4\,GHz, and was first detected in a deep search
with the Parkes telescope.  Subsequent observations with Parkes and the
Green Bank Telescope have confirmed this detection, and yield a period
derivative $\dot P = 2.02\times10^{-13}$. These spin parameters imply
a characteristic age $\tau_c = 4.8$\,kyr and a spin-down luminosity
$\dot E = 3.3\times10^{37}$\,\ergs, the latter value exceeded only by
the Crab pulsar among the rotation-powered pulsars known in our Galaxy.
The pulsar has an unusually steep radio spectrum in the 0.8--2.0\,GHz
range, with power law index $\approx 3.0$, and a narrow single-peaked
pulse profile with full-width at half maximum of $0.04P$.  We have
analyzed 350\,ks of archival {\em Chandra X-ray Observatory\/} High
Resolution Camera (HRC) data, and find a point-like source of luminosity
$\approx 3\times10^{-5} \dot E$, offset from the center of an elliptical
region of size $\approx 7'' \times 5''$ and luminosity $\approx 10^{-3}
\dot E$ within which likely lies the pulsar wind termination shock.
We have searched for X-ray pulsations in a 30\,ks HRC observation without
success, deriving a pulsed fraction upper limit for a sinusoidal pulse
shape of about 70\% of the pulsar flux.  We revisit the distance to \snr\
based on \HI\ and CO observations, arguing that it is $4.7 \pm 0.4$\,kpc.
We use existing X-ray and radio observations of the pulsar wind nebula,
along with the measured properties of its engine and a recent detection
of the supernova remnant shell, to argue that \snr\ and \psr\ are much
younger than $\tau_c$, and likely their true age is $\la 1000$\,yr.
In that case, the initial spin period of the pulsar was $\ga 55$\,ms.
\end{abstract}


\section{INTRODUCTION}\label{sec:intro} 

``Crab-like'' remnants of supernovae (SNe) display no readily observable
manifestation of an ejecta-driven blast wave; instead, the relativistic
wind of particles and magnetic fields released by a young pulsar is
confined by the ambient pressure giving rise to a non-thermal compact
nebula.  Such pulsar wind nebulae (PWNe), particularly with knowledge of
the spin parameters of their central engines, are magnificent laboratories
for the study of astrophysical shocks, the properties of pulsar winds,
and the spin evolution and birth parameters of neutron stars.  Eventually,
depending on explosion energy and environment, high velocity SN ejecta are
expected to generate a brightening shell surrounding the PWN, creating a
``composite'' supernova remnant (SNR).

\snr, known for 35 years \citep{adg+70,wa70} and long-classified
\citep{ww76,bs81} as one of about ten Crab-like remnants \citep{gre04},
is a bright and absorbed \citep[neutral hydrogen column density
$N_H=2\times10^{22}$\,cm$^{-2}$;][]{shp+01} centrally peaked X-ray and
polarized radio source.  \HI\ absorption measurements (Caswell et al.\
1975; Davelaar, Smith, \& Becker 1986)\nocite{cmr+75,dsb86} place the
system at $>4.4$\,kpc.  Its radio and X-ray luminosities are $\sim10$
and $\sim100$ times below those of the Crab Nebula, respectively
\citep{hb87}.  A break in the observed spectrum of \snr\ was originally
reported at a frequency $\nu_b \sim 50$\,GHz \citep[e.g.,][]{srh+89},
although more recently it has been suggested that $\nu_b \ga 500$\,GHz
\citep*{gt98,bwd01,bnc01}.

X-ray observations in the modern era have advanced our knowledge of
\snr\ enormously.  At the center of the system is a bright compact
source \citep[$\approx 2''$ in radius;][]{scs+00} that may include the
pulsar and delineate its wind termination shock.  Surrounding this is the
previously known PWN ($\approx 40''$ in radius), which is more compact
than the radio nebula, and with a spectrum that steepens radially from
the center \citep[e.g.,][]{shp+01,wbb+01,bwd01}.  But most remarkably,
surrounding the PWN is a low surface-brightness non-thermal halo,
$\approx 150''$ in radius, with no radio counterpart \citep{scs+00},
and apparently with a spatially invariant photon index \citep{ms05}.
The very deep {\em Chandra}\ image presented by \citet{ms05} beautifully
demonstrates evidence for limb brightening in this halo.  It now appears
likely that a portion of the emission observed in the halo originates
in the long sought SNR shell, while it is possible that dust scattering
also contributes significantly to the observed emission \citep{bscb05},
especially within about $80''$ of the center.  While unusual, such a
non-thermal shell is not unprecedented \citep[e.g.,][] {kpg+95,sgd+99},
and could be due to relativistic electrons accelerated at the SN
ejecta--ambient gas shock front.  \snr\ has also been detected in
hard X-rays by {\em INTEGRAL\/}, with a 20--100~keV flux of $\sim
3$\,millicrabs and a flatter spectrum than that of the Crab Nebula
\citep{bbb+04}.

Meanwhile, the pulsar powering the PWN at the center of the
composite SNR~\snr\ has remained undetected, despite searches
in the radio \citep[e.g.,][]{kmj+96,cpkm02a} and X-ray bands
\citep{scs+00,shp+01,pm02}.  We have now, in the course of extensive deep
radio searches of PWNe \citep[e.g.,][]{clb+02,cmg+02}, discovered \psr,
the central engine of \snr, which we report on here.  After submitting
this paper, we became aware of an independent discovery of this pulsar
by \citet{gmga05}, whose results on the pulsed signal from this source
largely agree with ours.

We begin by describing the pulsar search at radio wavelengths
(\S~\ref{sec:r-pulse}) and its confirmation and follow-up observations
(\S~\ref{sec:r-conf}).  We then present an X-ray imaging analysis of \snr\
(\S~\ref{sec:x-image}) and search for pulsations (\S~\ref{sec:x-pulse}).
In \S~\ref{sec:dist} we obtain the distance to the system, after which
we discuss the wind termination shock region, nebular magnetic field
(\S~\ref{sec:shock}), and age (\S~\ref{sec:age}) of \snr.  We compare
the properties of \psr\ and its SNR to those of SNR~3C~58 and its pulsar
in \S~\ref{sec:3c58}, and in \S~\ref{sec:r-spectra} we remark on the
radio spectrum of \psr, putting it in the context of those of other
young pulsars.  We summarize our work in \S~\ref{sec:sum}.

\section{OBSERVATIONS AND ANALYSIS}\label{sec:obs}

\subsection{Radio Pulsar Search}\label{sec:r-pulse}

We observed the position R.A. (J2000) $18^{\rm h}33^{\rm m}33\fs8$,
Decl. (J2000) $-10\arcdeg34'10''$, centered on \snr, at the Parkes
64-m radio telescope on 2002 January 6 for a full transit of 8.0\,hr
with the central beam of the multibeam receiver system \citep{swb+96}.
The observation was centered at a frequency of 1374\,MHz, with a bandwidth
of 288\,MHz for each of two polarizations split into 96 contiguous
frequency channels.  Total-power signals were detected and sampled with
1-bit precision every 0.8\,ms, with a total of 36 million time samples
for each of 96 frequency channels recorded to magnetic tape for analysis.

We analyzed the data using standard pulsar search techniques
\citep[see, e.g.,][]{lk05} implemented in the PRESTO software package
\citep*{ran01,rem02}.  While the \citet{cl02} model for the Galactic
distribution of free electrons predicts a dispersion measure (DM) $>
282$\,\ccpc\ for the distance to \snr\ of $>4.4$\,kpc, such predictions
can be inaccurate \citep[e.g.,][]{hcg+01,csl+02}, and we searched the
DM range 0--2674\,\ccpc, up to twice the maximum expected Galactic
value in this direction.  We first analyzed the non-dispersed (i.e.,
$\mbox{DM}=0$\,\ccpc) data to identify strong interference signals that
we subsequently masked (in the time, radio frequency, or fluctuation
frequency domains).  We de-dispersed, barycentered, and searched the
data for each of 524 trial DMs while maintaining close to optimal
time resolution throughout the DM range searched.  A pulsar with very
large spin period derivative, $\dot P$, may change its spin frequency
perceptibly within such a long observation, diminishing the sensitivity
unless a number of trial acceleration values are used to generate multiple
time series to be searched: e.g., the Crab pulsar's fundamental would move
by 0.3 Fourier bins within our observation, and a young pulsar in a binary
system could be affected even more by orbital Doppler accelerations.
For this reason we used Fourier-domain acceleration searches to cover
a range of $\dot P$ that would cause a signal to drift by up to $\pm
16$ Fourier bins for the highest of 1, 2, 4, 8, and 16 harmonics.
We summed these harmonics incoherently to retain sensitivity to narrow
pulse profiles.

\begin{figure}
\epsscale{0.90}
\plotone{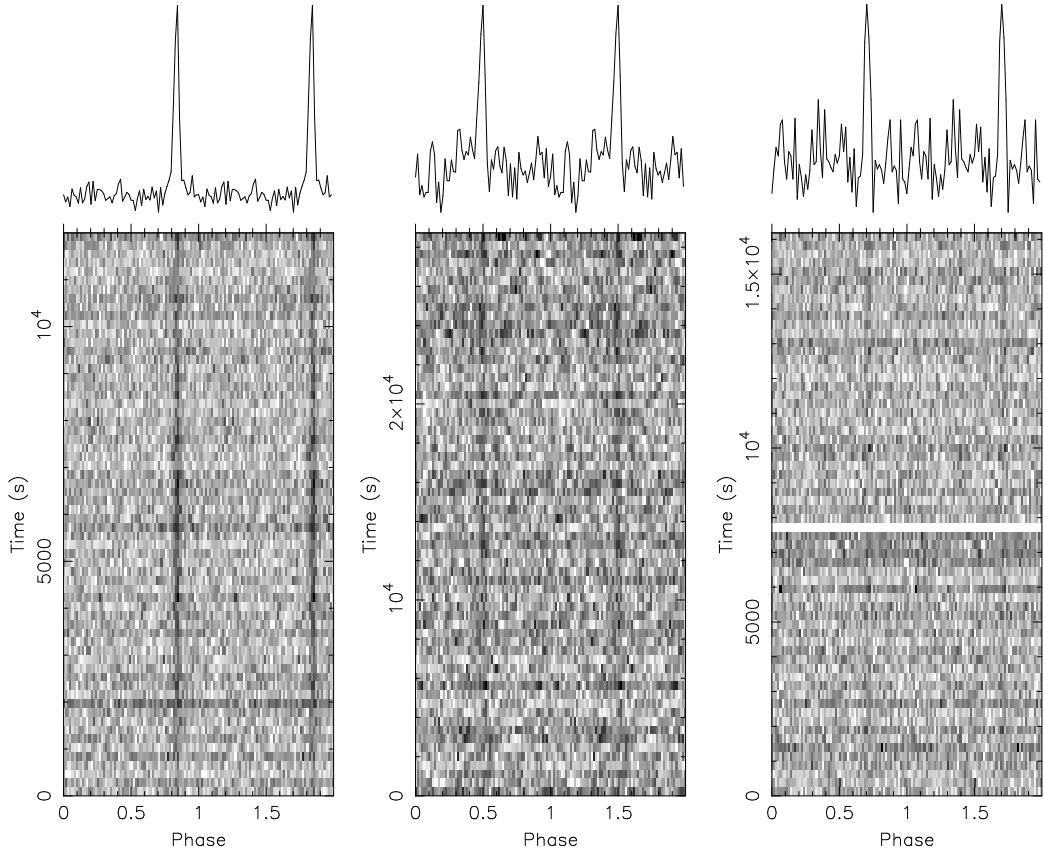}
\caption{ \label{fig:profs}
Radio pulse profiles of \psr.  Central frequencies increase from left
to right: 820\,MHz (GBT), 1374\,MHz (Parkes; discovery observation),
and 1850\,MHz (GBT).  In each plot, the linear grey-scale represents
pulsar strength as a function of rotational phase and time (increasing
from bottom), with the total observation split into and displayed
as 64 sub-integrations.  At top is shown the summed pulse profile
for the full observation.  Two pulse periods are shown in all cases.
The 1850\,MHz profile was obtained from a bandwidth of 400\,MHz (the
pulsar was not detected in the upper 200\,MHz of bandwidth recorded with
SPIGOT; see Table~\ref{tab:obs}), and the white band indicates removal
of data due to interference.  There is evidence of apparent interstellar
scintillation in the time-varying flux densities of the 820\,MHz and
possibly 1374\,MHz data sets (and the nominal flux densities for separate
820\,MHz and 1374\,MHz observations differ by $\sim 10\%$--$30\%$; see
Table~\ref{tab:obs}).  The observed pulse full-width at half-maximum is
3.0\,ms ($0.05P$) at 1374\,MHz and 2.5\,ms at the other two frequencies.
Smearing due to differential dispersive propagation within one frequency
channel is 1.3\,ms for the 820\,MHz data and 1.6\,ms for the 1374\,MHz
data (which, additionally, have a sample interval of 0.8\,ms).  Therefore,
the intrinsic pulse FWHM is very nearly $0.04P$ at all frequencies. }
\end{figure}

We detected a clear pulsar signal at $\mbox{DM} \approx 170$\,\ccpc\
with barycentric period $P = 61.84$\,ms, $\dot P < 5 \times 10^{-13}$
and a narrow profile with full-width at half maximum of $0.05P$
(Fig.~\ref{fig:profs}, middle panel).

\subsection{Confirmation and Follow-up Observations}\label{sec:r-conf}

On 2005 April 14 we observed the pulsar with the 100-m Green Bank
Telescope (GBT) for 0.5\,hr at a central frequency of 1950\,MHz.
While the pulsar was not detected in an unbiased way in this short
observation, searching a range of periods for assumed values of $\dot P
\la 5 \times 10^{-13}$ yielded the most promising candidate for $\dot P
\sim 2 \times 10^{-13}$.  On 2005 April 19, we clearly re-detected the
pulsar at 1374\,MHz at Parkes with $P=61.86$\,ms, implying $\dot P =
2.02 \times 10^{-13}$.  The retrospective detection at GBT also implies
that its positional uncertainty from radio observations is about $\pm 3'$
(telescope beamwidth).

We have now observed the new pulsar at 820\,MHz, in addition to the
two frequencies described above (see Table~\ref{tab:obs}).  There is
some evidence for fluctuation in received flux density, likely due to
interstellar scintillation (see Fig.~\ref{fig:profs}).  The flux densities
listed in Table~\ref{tab:obs} are calibrated only approximately, and are
therefore subject to some future adjustment.  Nevertheless, we infer
that the spectrum of the pulsar is steep, with a weighted fit to the
values presented in Table~\ref{tab:obs} giving $\alpha = 3.0\pm0.3$
(throughout this paper, $S_\nu \propto \nu^{-\alpha}$).  The flux
density of $\approx 0.65$\,mJy at a frequency of 610\,MHz estimated by
\citet{gmga05} is consistent with this determination of $\alpha$.

\begin{deluxetable}{llcrrlcr}
\tablewidth{0pt}
\tablecaption{\label{tab:obs}Radio Observations of \psr. }
\tablecolumns{8}
\tablehead{
\colhead{Date}                           &
\colhead{Telescope}                      &
\colhead{Time}                           &
\colhead{Frequency}                      &
\colhead{Bandwidth}                      &
\colhead{Backend\tablenotemark{a}}       &
\colhead{$T_{\rm sys}$\tablenotemark{b}} &
\colhead{$S_{\rm PSR}$\tablenotemark{c}} \\
\colhead{(MJD)}     &
\colhead{}          &
\colhead{(hr)}      &
\colhead{(MHz)}     &
\colhead{(MHz)}     &
\colhead{}          &
\colhead{(K)}       &
\colhead{($\mu$Jy)}
}
\startdata
52280 & Parkes & 8.0 & 1374 & 288 & filter bank & 40 & 66      \\
53474 & GBT    & 0.5 & 1950 & 600 & SPIGOT      & 39 & \nodata \\
53479 & Parkes & 5.9 & 1374 & 288 & filter bank & 40 & 76      \\
53482 & GBT    & 3.3 &  820 &  48 & BCPM        & 83 & 277     \\
53486 & GBT    & 4.5 & 1950 & 600 & SPIGOT      & 39 & 13      \\
53502 & GBT    & 0.4 &  820 &  48 & BCPM        & 83 & 283     \\
53525 & GBT    & 1.0 &  820 &  48 & BCPM        & 83 & 181     \\
53557 & GBT    & 0.8 &  820 &  48 & BCPM        & 83 & 193     \\
53575 & GBT    & 2.8 &  820 &  48 & BCPM        & 83 & 171     \\
\enddata
\tablenotetext{a}{For details on observing equipment, see: filter bank
\citep{mlc+01}; SPIGOT \citep{rhs+05,kel+05}; BCPM \citep{bdz+97,csl+02}. }
\tablenotetext{b}{Includes contribution from Galaxy \citep[scaled
from the 408\,MHz map of][with spectral index of 2.6]{hssw82} and SNR
\citep[$\approx 6$\,Jy at 1\,GHz with flat spectrum;][]{srh+89}. }
\tablenotetext{c}{Integrated period-averaged pulse profile, compared
to the rms of the off-pulse region, scaled via the radiometer equation
using the observation bandwidth and time, and assuming the nominal
telescope gain and system temperature (previous column).  For SPIGOT,
we used only 1650--2050\,MHz data to calculate flux density (see caption
to Fig.~\ref{fig:profs}).  These estimates have an uncertainty of about
25\%, exclusive of variations due to interstellar scintillation.  }
\end{deluxetable}

The new pulsar has a magnetic field strength at its light cylinder of
$B_{\rm lc} = 1.4\times 10^5$\,G, the sixth highest value among pulsars
known in the Galactic disk, ex aequo with PSR~J2229+6114 \citep{hcg+01}.
Since the emission of ``giant'' radio pulses appears to be correlated with
$B_{\rm lc}$ \citep[e.g.,][]{cst+96,jr03}, we made a preliminary search
for such pulses in the long 820\,MHz data set (Table~\ref{tab:obs}),
but have not detected any.

We measured pulse times-of-arrival (TOAs) from the GBT and Parkes data and
fitted a simple spin-down model to the 2005 TOAs using TEMPO\footnote{See
http://pulsar.princeton.edu/tempo.}.  The results of the fit are shown
in Table~\ref{tab:parms}.  The position used for the pulsar was that
determined from \chandra\ observations (\S~\ref{sec:x-image}).  The $\dot
P$ measured from the phase-coherent solution at the present epoch matches
that derived from two period measurements separated by 3.5\,yr.

\begin{deluxetable}{ll}
\tablewidth{0pt}
\tablecaption{\label{tab:parms}Measured and Derived Parameters for \psr. }
\tablecolumns{2}
\tablehead{
\colhead{Parameter} &
\colhead{Value}
}
\startdata
Right ascension, R.A. (J2000)\dotfill & $18^{\rm h}33^{\rm m}33\fs57$         \\
Declination, Decl. (J2000)\dotfill    & $-10\arcdeg34'07\farcs5$              \\
Spin period, $P$ (s)\dotfill          & 0.061865672117(5)                     \\
Epoch (MJD)\dotfill                        & 53545.0                          \\
Period derivative, $\dot P$~\tablenotemark{a}\dotfill & $2.02025(3)\times 10^{-13}$         \\
Period derivative, $\dot P$~\tablenotemark{b}\dotfill & $(2.0206\pm0.0004)\times 10^{-13}$ \\ 
Dispersion measure, DM (\ccpc)\dotfill     & $169.5(1)$                       \\
Post-fit rms timing residuals (ms)\dotfill & 0.23                             \\
Data span (MJD)\dotfill                    & 53479--53575                     \\
Pulse full-width at half-maximum (ms)\dotfill           & $\approx 2.5$       \\
Flux density at 820\,MHz ($\mu$Jy)\dotfill              & $\approx 221$       \\
Flux density at 1374\,MHz, $S_{1400}$ ($\mu$Jy)\dotfill & $\approx  71$       \\
Flux density at 1850\,MHz ($\mu$Jy)\dotfill             & $\approx  13$       \\
%\colrule
Spin-down luminosity, $\dot E$ (\ergs)\dotfill          & $3.37\times10^{37}$ \\
Surface dipole magnetic field strength (gauss)~\tablenotemark{c}\dotfill  & $3.58\times 10^{12}$ \\
Characteristic age, $\tau_c$ (kyr)\dotfill              & 4.85                \\
Distance, $d$ (kpc)\dotfill                             & $4.7 \pm 0.4$       \\
Radio luminosity at 1.4\,GHz, $S_{1400} d^2$ (mJy\,kpc$^2$)\dotfill & $\approx 1.5$       \\
Radio spectral index, $\alpha$\dotfill                  & $\approx 3.0$       \\
\enddata
\tablecomments{Celestial coordinates are assumed to correspond to
the central peak of the X-ray PWN seen in \chandra\ HRC observations
(\S~\ref{sec:x-image}; Fig.~\ref{fig:hrc}).  Numbers in parentheses
are three times the nominal 1\,$\sigma$ TEMPO uncertainties in the
least-significant digits quoted. }
\tablenotetext{a}{This $\dot P$ was determined from the timing solution
based on data collected during the span listed. }
\tablenotetext{b}{This $\dot P$ was obtained from the TEMPO-derived $P$
for MJD 53545.0 and the period measured from the discovery observation
on MJD 52280. }
\tablenotetext{c}{$B_p \equiv 3.2\times10^{19} (P \dot P)^{1/2}$\,G
\cite[see, e.g.,][]{mt77}. }
\end{deluxetable}

\subsection{X-ray Imaging of \snr}\label{sec:x-image}

\begin{figure}
\plotone{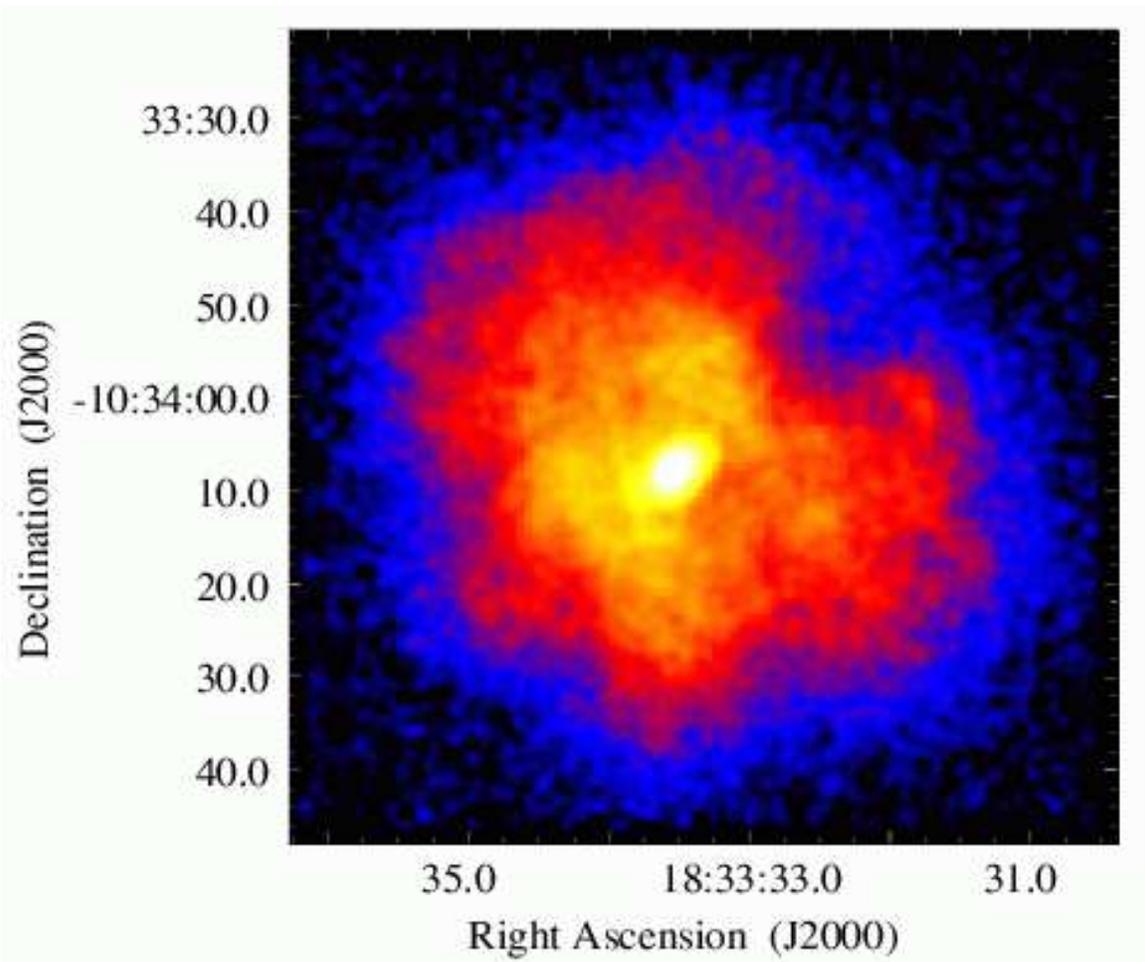}
\caption{ \label{fig:acis} 
False-color \chandra\ ACIS-S image of the PWN in SNR~\snr\ based on 58\,ks
of data, showing the elongated core embedded in the more diffuse nebula.
The SNR shell lies beyond the edge of this image.  The image has been
smoothed by two pixels ($0\farcs98$), and the intensity levels increase
logarithmically from a surface brightness of 9\,cts\,arcsec$^{-2}$ in the
outer regions (blue) to a value of 1640\,cts\,arcsec$^{-2}$ in the core
(white).  See \S~\ref{sec:x-image} for further details about the image.  }
\end{figure}

\snr\ is used as a calibration target for the {\em Chandra X-ray
Observatory}, and multiple observations exist using a variety of detector
modes and aimpoint positions.  For a set of six observations carried
out with the Advanced CCD Imaging Spectrometer (ACIS), with the target
placed within $1'$ of the optical axis of the telescope and falling on
CCD S3, we used the CIAO\footnote{See http://cxc.harvard.edu/ciao.}
task {\em merge\_all\/} to generate a combined image of the PWN
(Fig.~\ref{fig:acis}).  The integrated exposure for the image is 58\,ks,
and we have applied a Gaussian smoothing of two pixels ($0\farcs98$).
The image shows an extended core, identified in earlier \chandra\
observations \citep{scs+00,shp+01}, surrounded by a broader nebula
with a radius of $\approx 40''$.  There is considerable structure
in the outer nebula region, including filamentary features and a
distinct low-surface-brightness ``notch'' in the outer northwest
of the PWN \citep[see also][]{shp+01,ms05}.  We split the image
in Figure~\ref{fig:acis} into two energy bands (0.3--6.0\,keV and
6.0--10.0\,keV) and computed radial surface brightness profiles: the
profile in the hard band is slightly steeper than that in the soft
band, although X-ray emission extends in both bands to roughly the same
outer radius.

\begin{figure}
\plotone{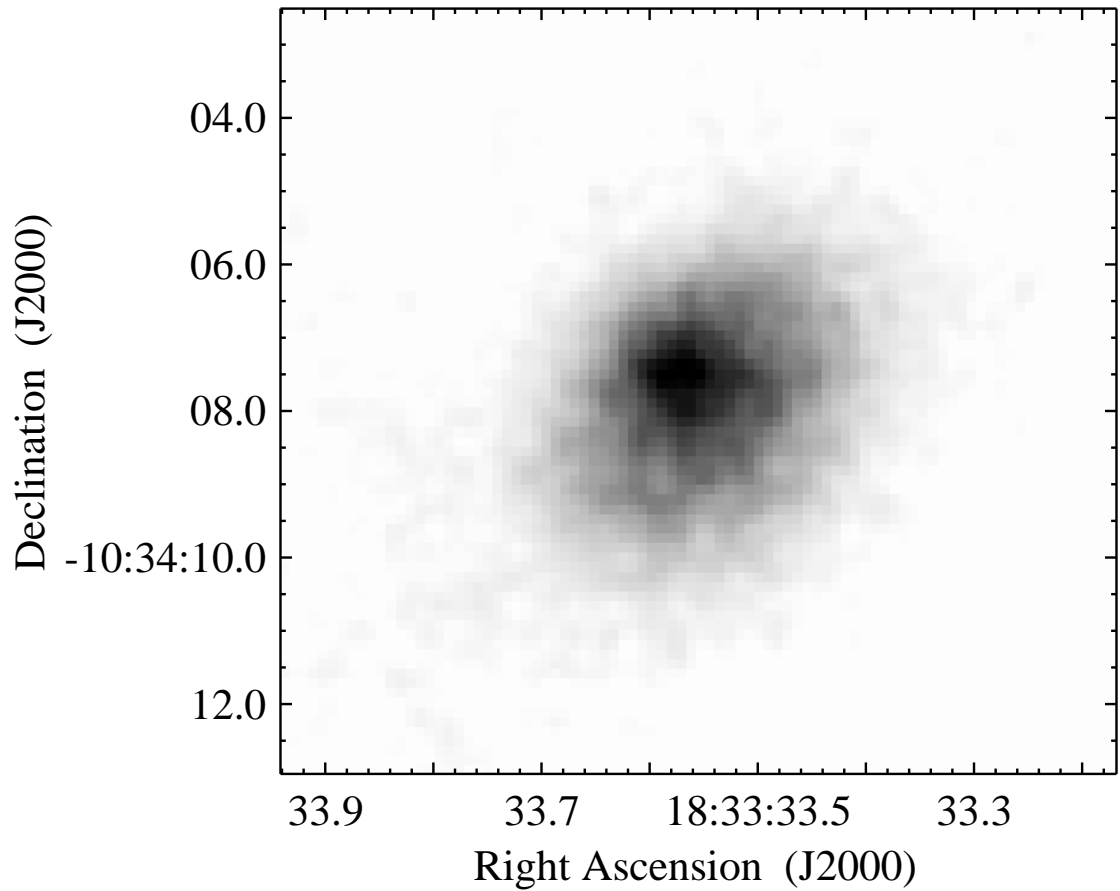}
\caption{ \label{fig:hrc} 
Grey-scale \chandra\ HRC image of the central core in \snr\
based on 351\,ks of data from multiple observations (compare to
Fig.~\ref{fig:acis}).  The image has been smoothed by two pixels
($0\farcs26$), and any residual registration error/blurring is at the
level of $<0\farcs3$.  A compact source, presumably corresponding to
\psr, is surrounded by an elongated emission region.  The grey-scale
intensity increases linearly from 520\,cts\,arcsec$^{-2}$ at the
outer edge of the core (e.g., near R.A. $18^{\rm h}33^{\rm m}33\fs5$,
Decl. $-10\arcdeg34'05\farcs5$) to 2500\,cts\,arcsec$^{-2}$ at the
position of the compact source.  See \S\S~\ref{sec:x-image} and
\ref{sec:x-pulse} for more details. }
\end{figure}

To further investigate the compact core, we created a merged image
from 21 observations with the \chandra\ High Resolution Camera (HRC),
using data from both the imaging (HRC-I) and spectroscopy (HRC-S) cameras
(the latter used here without transmission gratings).  The emission-line
star SS~397 was identified in the field of each observation, and its
2MASS position (R.A. [J2000] $18^{\rm h}33^{\rm m}27\fs70$, Decl. [J2000]
$-10\arcdeg35'23\farcs0$) was used to register separately each individual
observation in order to improve the overall astrometry.  The resulting
merged image of the core region, corresponding to a total integration
of 351\,ks, is shown in Figure~\ref{fig:hrc}.  With a typical 2MASS
positional uncertainty of $\la 0\farcs2$ and determination of the X-ray
centroid to better than an HRC pixel, we estimate that any residual
blurring/registration error in this image is at the level of $<0\farcs3$.
A comparison of Figure~\ref{fig:hrc} with Figure~4 of \citet{shp+01}
reveals significant differences in morphology, plus an offset of $\sim
1\farcs5$ in the position of peak emission.  \citet{shp+01} did not
discuss any efforts to correctly register each exposure, and it is
possible that their image has been affected by slight shifts between
observations.

Immediately evident in Figure~\ref{fig:hrc} is a compact source surrounded
by an elliptical emission region of size $\approx 7'' \times 5''$
(in a very saturated image, this core appears to be about 50\% larger
in each dimension; see, e.g., Fig.~\ref{fig:acis}).  This is to be
compared with the 90\% encircled-energy radius of $\approx 1''$ for HRC.
The point-like source is located at R.A. (J2000) $18^{\rm h}33^{\rm
m}33\fs57$, Decl. (J2000) $-10\arcdeg34'07\farcs5$ and offset from the
center of the ellipse, possibly indicating a Doppler-boosted toroidal
structure surrounding a pulsar, as is observed for the Crab and other
pulsars imaged with \chandra\ \citep[e.g.,][]{nr04,hrr+04,shvm04}.
In this scenario, the inferred pulsar spin axis projected on the sky is
perpendicular to the major axis of the ellipse, at a position angle of
$\sim 40\,\arcdeg$ north through east \citep[similar to the position
angle first suggested by][for the central source in \snr]{fhm+88}.
The major axis of the ellipse is aligned toward the notch of reduced
emission on the northwest rim of the PWN \citep[Fig.~\ref{fig:acis};
see also][]{ms05}.  We note the similarity with the ``bays'' observed
in the Crab Nebula, which are roughly aligned with the torus surrounding
its pulsar \citep*[e.g.,][]{fms92}.

In order to estimate the flux of what is presumed to be the pulsar,
we used the CIAO package Sherpa to perform spatial modeling in which we
fit the core region of \snr\ (Fig.~\ref{fig:hrc}) to a model consisting
of a point source (convolved with the PSF, calculated using the CIAO
task {\em mkpsf\/} for an energy of 1\,keV), a 2-D Gaussian, and a
constant background.  While the extended emission is clearly not a 2-D
Gaussian (we expect perhaps a Doppler-brightened torus superimposed on
more uniform emission), this is the most suitable model available in
Sherpa and provides a good fit.  As expected, the best-fit values for the
point-source position and Gaussian centroid are very different, limiting
the point source flux assigned to the Gaussian component.  In any case,
there is a systematic uncertainty in this flux determination in excess
of the formal fit uncertainty of 10\%, and a more realistic geometrical
modeling is required to improve significantly on this result.  In the
energy range 0.5--10\,keV, the resulting count rate for the pulsar is
$1.5\times10^{-3}$\,s$^{-1}$, corresponding to an unabsorbed flux of
about $2.8\times10^{-13}$\,ergs\,cm$^{-2}$\,s$^{-1}$ for a spectral
index of $\alpha_X = 0.5$, assuming $N_H = 2.2\times10^{22}$\,cm$^{-2}$
\citep{shp+01}, or a luminosity of about $9\times10^{32}$\,\ergs\
at a distance of 5\,kpc (see \S~\ref{sec:dist}).  The assumed $N_H$
may be somewhat in error, since simple fits to the data yield values
of $N_H$ that differ by as much as 20\% depending on radial position
within the nebula.  However, such a variance in $N_H$ introduces
a variation of only 20\% on the point source flux, smaller than the
uncertainty arising from the difficulties involved in spatial modeling.
Through such an absorbing column and at a distance of 5\,kpc, thermal
radiation from a 10\,km-radius neutron star with a surface temperature
of 1.4\,MK (appropriate for standard cooling to an age $\sim 1000$\,yr;
see \S~\ref{sec:age}) would contribute an HRC count rate of only $\sim
10\%$ that observed, and we do not consider it further here.

The entire elliptical region of X-ray emission shown in
Figure~\ref{fig:hrc}, including the compact component, has a 0.5--10\,keV
count rate of $6.1\times10^{-2}$\,s$^{-1}$.  Assuming $\alpha_X = 1.0$
and $N_H = 2.2\times10^{22}$\,cm$^{-2}$, this corresponds to an unabsorbed
0.5--10\,keV flux of $1\times10^{-11}$\,ergs\,cm$^{-2}$\,s$^{-1}$ or a
luminosity at 5\,kpc of $3\times10^{34}$\,\ergs.

\subsection{Search for X-ray Pulsations}\label{sec:x-pulse}

\snr\ was observed for 30\,ks with the \chandra\ HRC-S on 2002 November 12
(ObsID 2755), with the detector operated in timing mode.  In this mode,
event arrival times are adjusted to compensate for a wiring error,
allowing photons to be time-tagged to a resolution of $16\,\mu$s.
We corrected these event times to the Solar System barycenter
using the CIAO task {\em axbary\/} and the nominal pulsar position
(\S~\ref{sec:x-image}).  We extracted 558 events from a circular region of
radius $1''$, centered on the position of the compact source identified
in the merged HRC image (Fig.~\ref{fig:hrc}).  We searched these events
for pulsations around the spin frequency predicted by the radio ephemeris
(Table~\ref{tab:parms}), covering approximately 3 independent Fourier
spacings ($10^{-4}$\,Hz).  We used the Discrete Fourier Transform,
which is optimally sensitive to sinusoidal pulsations, and the Bayesian
technique of \citet{gl92b}, which is sensitive to pulsations with a more
complex but unknown shape.  In neither case did we find any evidence of
X-ray pulsations.  We estimate a 99\% confidence-level upper limit on
the pulsed fraction for sinusoidal pulsations of 21\%, using the method
described by \citet*{rgs02}.  If the pulsar has a significantly narrower
X-ray pulse profile, as in the radio, the pulsed fraction would be
significantly less than 21\% (with the exact value depending sensitively
on the pulse shape).  This limit pertains to pulsations from the total
observed flux within the extraction region and does not apply to the
pulsar's flux alone.  Since the pulsar likely contributes $\la 30\%$
of the counts in the region considered, our upper limit represents a
significant fraction ($\sim 70\%$) of the total signal from the pulsar,
and hence is not especially constraining.

\section{DISCUSSION}\label{sec:disc}

The $P$ and $\dot P$ of the newly discovered pulsar imply a characteristic
age $\tau_c \equiv P/(2 \dot P) = 4.8$\,kyr and spin-down luminosity
$\dot E \equiv 4 \pi^2 I \dot P P^{-3} = 3.3 \times 10^{37}$\,\ergs,
where the neutron star moment of inertia is $I \equiv 10^{45}$\,g\,cm$^2$.
This $\dot E$ is remarkably similar to that predicted by \citet{scs+00}
for the pulsar in \snr\ based on the X-ray energetics of the PWN, and is
second only to that of the Crab among known Galactic rotation-powered
pulsars.  Still, until X-ray pulsations are detected from the core of
the PWN with $P=61.8$\,ms or until the positional uncertainty of the
pulsar is reduced to a few arcsec, it could be argued that it may not
be the counterpart of the PWN.  However, if the newly discovered pulsar
were not associated with \snr, we would certainly expect to detect it
in the deep \chandra\ observations of the field, given its relatively
small inferred distance (\S~\ref{sec:dist}) and large $\dot E$, and no
such potential counterpart exists \citep[e.g.,][]{ms05}.  Therefore,
this is clearly the neutron star responsible for the PWN in \snr, which
we designate \psr\ based on the position of the central compact source
detected with \chandra\ (Fig.~\ref{fig:hrc}; see also \S~\ref{sec:x-image}
and Table~\ref{tab:parms}).

\subsection{Distance to \snr\ and \psr}\label{sec:dist}

To derive an upper limit on the distance to \snr, we consider the bright
extragalactic source PMN~J1832--1035, which lies just $18'$ from \snr.
An \HI\ absorption spectrum of PMN~J1832--1035, using data taken from
the Very Large Array Galactic Plane Survey \citep{tsd+02}, shows a
strong absorption feature at a local standard of rest (LSR) velocity
of +76\,\kms, but which is not seen toward \snr\ \citep{cmr+75,dsb86}.
The CO survey of \citet*{dht01} shows a bright coherent region of
molecular material at this velocity, present in the direction of both
PMN~J1832--1035 and \snr.  Thus, it is highly likely that there is
cold neutral gas at an LSR velocity of +76\,\kms\ toward both sources.
The lack of \HI\ absorption against \snr\ at this velocity implies that
the SNR is in front of this cloud.

We next compare the emission spectrum of this CO to the \HI\ emission
survey of \citet{hb97}.  In the direction of PMN~J1832--1035, the \HI\
emission shows a narrow feature of reduced emission in the spectrum at a
velocity of +76\,\kms.  We interpret this feature as \HI\ self-absorption,
indicating that cold gas on the near side of the tangent point is
absorbing \HI\ emission produced on the far side (Liszt, Burton, \& Bania
1981; Jackson et al.\ 2002\nocite{lbb81,jbs+02}).  Using the Galactic
rotation curve of \citet*{fbs89} and assuming a Galactocentric radius
for the Sun of 8.5\,kpc, the distance to this cloud and hence an upper
limit on the distance to \snr\ is 4.9\,kpc.  Combined with previous \HI\
measurements that put a lower limit on the SNR's systemic LSR velocity of
65\,\kms\ \citep{dsb86}, corresponding to a distance limit of 4.4\,kpc,
and adopting an uncertainty in the systemic velocity of \HI\ clouds
of $\pm7$\,\kms\ \citep{sra+82,bc84}, we conclude that the kinematic
distance to SNR~\snr\ is $4.7 \pm 0.4$\,kpc.

The \cite{cl02} model yields a distance $d = 3.3$\,kpc for \psr, which is
clearly an underestimation of the true distance, but supports a relatively
small value.  \citet{shp+01} have estimated $4.1 < d < 5.2$\,kpc from
a calibration to a range of X-ray-derived $N_H$ values.  Hereafter we
scale the distance to \psr\ and \snr\ in terms of $d_5 \equiv d/$(5\,kpc).

The X-ray spectra of the various components of \snr\ imply a
foreground absorbing column $N_H \approx 2\times10^{22}$\,cm$^{-2}$
\citep{wbb+01,shp+01}.  Comparison with the DM given in
Table~\ref{tab:parms} implies a ratio $N_H/{\rm DM} \approx 40$.  This is
substantially higher than seen toward all but $\sim3$ other pulsars
\citep[see discussion in][]{gvc+04}.  Gaensler et al.\ argue that these
high ratios indicate that a source is behind a substantial amount of
molecular material.  Indeed, the CO data of \citet{dht01} indicate at
least three CO clouds at LSR velocities between 0 and 75~\kms, which
presumably contribute to the high observed neutral column.

\subsection{Relativistic Wind Termination Shock and Nebular Magnetic
Field}\label{sec:shock}

Knowing $\dot E$ for \psr, we can estimate the termination shock radius,
$r_s$, at which the pulsar wind pressure, $\dot E/(4 \pi r_s^2 \eta c)$,
balances the interior pressure of the nebula, $P_n$, where $\eta$ is
the fractional solid angle covered by the pulsar wind.  Assuming nebular
equipartition between particles and fields, $P_n = B_n^2/(4 \pi)$, and
we have $\theta_s = 0\farcs46\,\eta^{-1/2} d_5^{-1} B_{\rm mG}^{-1}$,
where $B_{\rm mG}$ is the nebular field in mG and $\theta_s = r_s / d$.
The field derived from equipartition arguments \citep[e.g.,][]{pac70}
for the whole PWN is $B_n \approx 0.3$\,mG \citep[where we have used
a spectral index of $\alpha_r = 0$ between 100\,MHz and 500\,GHz;
see][and references therein]{bnc01}, implying $\theta_s \approx
1\farcs6\,\eta^{-1/2}$.  This can be compared with Figure~\ref{fig:hrc},
from which we infer that $\theta_s \la 2\farcs5$ (note however that we do
not actually observe a limb-brightened ring-like structure, as is the case
in some PWNe, and it is presumably also possible that $\theta_s \la 1''$).
The approximate match between estimated and inferred shock radii suggests
that the nebular magnetic field strength is indeed relatively large,
as estimated from equipartition arguments.

For such a relatively high field, the synchrotron lifetime of
relativistic electrons that emit 2\,keV photons is only $\sim 20$\,yr.
Assuming a nebular flow velocity profile of $V(r) = (c/3) (r_s/r)^2$
\citep{kc84a}, these electrons would travel $\sim 12''$ before radiative
losses become important.  (The equivalent nebular flow timescale to
the edge of the PWN is $\ga 500$\,yr.)  According to \citet{ms05},
the X-ray spectrum steepens significantly beyond its slope close to
the pulsar at a radius $\ga 10''$--$15''$ (which also appears to be the
approximate angular scale of the relatively bright X-ray ``plateau'' seen
in Fig.~\ref{fig:acis}).  Additionally, it appears that the normalized
X-ray and radio surface brightness profiles begin to diverge, with the
nebula becoming effectively smaller at X-ray energies, at $\ga 10''$
\citep[][see also \S~\ref{sec:x-image}]{bwd01,dw04}.  Overall, it thus
appears that a relatively high nebular field strength is compatible with
many of the observations.

A break in the spectrum of \snr\ is inferred at $\nu_b \ga 500$\,GHz
\citep[$\lambda_b \la 1$\,mm;][]{gt98,bwd01,bnc01}.  However, for an
age $\ga 500$\,yr and $B_n \approx 0.25$\,mG, one expects a break in the
spectrum due to synchrotron cooling at $\la 1\,\mu$m.  The sub-millimeter
break must then be due to something else, presumably reflecting the
complex nature (or variability) of the spectrum injected by the pulsar
or of its conversion to detectable electromagnetic radiation.

\subsection{The Evolutionary State and Age of \snr}\label{sec:age}

For spin evolution with constant magnetic moment, the age of a pulsar
is $\tau = P \dot P^{-1} (n-1)^{-1} [1 - (P_0/P)^{n-1}]$; the index $n$
is 3 for dipole braking \citep[in the few cases measured, $1.5 \la
n \la 3$; e.g.,][and references therein]{ckl+00}.  Further assuming
that the initial spin period $P_0 \ll P$, reduces the age to $\tau_c$,
or 4.8\,kyr in the case of \psr.  Hereafter we scale the age of the
pulsar/SNR system by $\tau_5 \equiv \tau/$(5\,kyr), and now show from
multiple lines of evidence that in fact $\tau_5 \ll 1$.

The scale of structures in \snr\ is $R = 1\,d_5 \theta_{40}$\,pc,
where we scale angular sizes by $\theta_{40} \equiv \theta/$($40''$).
Corresponding velocities are $V = 200\,d_5 \tau_5^{-1} \Delta
\theta_{40}$\,\kms, where $\Delta \theta_{40}$ is a change of angular
position in units of $40''$.  For the PWN, $\theta_{40} \approx 1$,
and for the SNR shell, $\theta_{40} \approx 150/40$; thus, $R_{\rm PWN}
\approx 1\,d_5$\,pc and $R_{\rm SNR} \approx 3.6\,d_5$\,pc.  Both of
these are small, similar to the corresponding sizes in G11.2--0.3,
a 1600 year-old composite SNR \citep{krv+01,che05}.

\psr\ is located very near the geometrical center of the \snr\ shell.
Using the image presented by \cite{ms05}, we estimate that the pulsar's
offset from the center is $\Delta \theta \la 5''$.  If the center
corresponds to the SN site, the implied projected space velocity is
$V_{\rm PSR} \la 25\,d_5 \tau_5^{-1}$\,\kms.  Unless $\tau_5 \ll 1$, this
is a very small velocity for a young pulsar \citep[e.g.,][according to
whom the probability of $V_{\rm PSR} < 25$\,\kms\ is 0.45\%]{ll94,hllk05}.
This reasoning was used also by \cite{krv+01} to argue that $\tau \ll
\tau_c$ for the pulsar in G11.2--0.3.

The shell in \snr\ is very faint in X-rays \citep{ms05} and has not yet
been detected at radio wavelengths \citep{scs+00}.  This suggests that
the system is only slightly more evolved than the Crab, i.e., the shell
is only now becoming visible and the SNR is on its way to becoming
a clear-cut composite like G11.2--0.3 (in detail, such an argument
requires of course an understanding of the circumstellar environments and
explosion dynamics).  In any case, for a uniform hydrogen ambient medium
of density $n_0$\,cm$^{-3}$, the SNR has swept up only $M_{\rm sw} \sim
5\,n_0 d_5^3$\,\msun\ and thus should be in nearly free expansion.  For an
ejected mass of $M_{\rm ej} = 10\,M_{\rm 10}$\,\msun\ and SN explosion
kinetic energy of $10^{51}\,E_{51}$\,ergs, the expansion velocity would
be $3200\,(E_{51}/M_{\rm 10})^{1/2}$\,\kms, implying $\tau_5 = 0.25\,d_5
(M_{\rm 10}/E_{51})^{1/2}$.  Since a freely expanding SNR would in reality
have a steep density profile to its ejecta, these estimates of  expansion
velocity and age are likely to be lower and upper limits, respectively.

Alternatively, let us consider that the expansion has entered
the adiabatic phase.  Using the Sedov solution, $R_{\rm SNR} =
0.31\,(E_{51}/n_0)^{1/5}t^{2/5}_{\rm yr}$\,pc.  If $\tau_5 = 1$,
then $n_0/E_{51} = 120\,d_5^{-5}$.  Such a high ambient density is
clearly inconsistent with the low levels of observed X-ray emission.
If conversely we adopt typical values of $n_0/E_{51} \la 1$, then
$\tau_5 \la 0.1\,d_5^{5/2}$.  But in this case the argument is not
self-consistent, because then we would have $M_{\rm sw} < M_{\rm ej}$.
The SNR is therefore still in nearly free expansion, for which the
arguments in the above paragraph apply\footnote{If $M_{\rm ej} \ll
10$\,\msun, as can occur, for instance, in type Ib/Ic SNe, then it does
not follow directly from these simple arguments that $M_{\rm sw} < M_{\rm
ej}$.  However, in a detailed spectroscopic and spatial analysis of X-ray
data on \snr, \cite{bscb05} suggest that $M_{\rm sw} \la 0.5$\,\msun,
and that in fact the remnant has not yet entered the adiabatic phase.}.

Consideration of PWN evolution arguments also suggests $\tau_5 \ll 1$.
In general, the ratio $R_{\rm PWN}/R_{\rm SNR}$ is initially small as
the freely expanding ejecta move faster than the PWN expansion rate
of $\sim 1000$\,\kms.  However, the PWN expansion accelerates with the
energy input by the pulsar and this ratio increases quickly.  After the
SNR ejecta decelerate and the reverse shock begins to compress the PWN,
the ratio decreases.  Since the PWN in \snr\ is approximately circular,
is symmetrically placed within the SNR shell, and has the pulsar at its
center, it is likely to be in its early pre-reverse-shock evolutionary
stage.  For the observed ratio $R_{\rm PWN}/R_{\rm SNR} \approx 0.25$,
the models of \citet*{bcf01} are indicative of a very small age.
Alternatively, considering simply a constant PWN expansion velocity
of $1000\,V_{1000}$\,\kms\ \citep[for comparison, the Crab Nebula is
expanding at $\sim1500$\,\kms;][and references therein]{bkhw91}, the
observed PWN size implies $\tau_5 = 0.2\,d_5 V_{1000}^{-1}$.

The PWN energetics also provide useful insight into the age of \snr.
According to \citet{che05}, the ratio $E_{\rm int}/(\dot E \tau)$
separates ``young'' (ratio $< 1$) from older PWNe, where $E_{\rm
int}$ is the internal energy of the PWN.  A lower limit on $E_{\rm
int}$ can be calculated from synchrotron emission, equivalent to
approximate equipartition between particles and fields, which is
roughly consistent with observations of PWNe.  We use equation~(46)
of \citet{che05} to calculate that for \snr, $E_{\rm min} \sim
2.9\times10^{47}\,d_5^{17/7}$\,ergs.  Here we have used $p_1 = 1 + 2
\alpha_r \approx 1.04$ \citep{srh+89} and $p_2 = 1 + 2 \alpha_X \approx
3$ \citep{shp+01}, as well as $\nu_b = 500$\,GHz and $L_{\nu b} \approx
1.2\times10^{23}\,d_5^2$\,\ergs\,Hz$^{-1}$, where $p_1$ and $p_2$ are the
energy indices for radio- and X-ray-generating particles, respectively,
and $L_{\nu b}$ is the spectral luminosity at $\nu_b$.  With the measured
$\dot E$ for \psr, we have $\dot E \tau = 5.3\times10^{48}\,\tau_5$\,ergs,
which is larger than $E_{\rm int}$ (typically within a factor of a few
of $E_{\rm min}$) for any reasonable age.  This then implies $\tau \ll
\tau_c$ \citep[see][]{che05}.

These various arguments all lead to the conclusion that for \psr\
and \snr, $\tau_5 \ll 1$, and maybe $\tau \la 1000$\,yr \citep[a
similar conclusion was reached independently by][who in an analysis
of X-ray data on \snr\ obtain an age range of 200--1000\,yr]{bscb05}.
In turn this suggests that $P_0 \ga 55$\,ms essentially independent
of braking index.  Measuring $n$ would of course be very interesting
and would further constrain the rotational history of J1833--1034, but
this is often impossible.  For example, for PSR~J0205+6449 in SNR 3C~58,
the rotational instability of the neutron star prevents $n$ from being
determined \citep{rck+04}.

While the age inferred here would make \psr\ among the 2--3 youngest known
neutron stars in the Galaxy, it is highly unlikely that its progenitor SN
was observed as an historical event \citep[and indeed no SN is recorded
in this direction in the compilation of][]{sg02b}. The observed hydrogen
column density corresponds to a visual extinction of 10--11 magnitudes.
The peak visual magnitude of the supernova might thus have been only
5--6 (for an absolute magnitude $M_V=-18.5$), which is likely to have
gone unnoticed.

\subsection{Comparison of \snr, 3C~58 and their Pulsars}\label{sec:3c58}

The measured spin parameters of \psr\ are very similar to those of
PSR~J0205+6449 \citep{mss+02}, the central engine in 3C~58 \citep[another
famous SNR long-classified as Crab-like, from which faint thermal
emission has now been detected; see][]{bwm+01,shvm04}.  It is therefore
of interest to compare the properties of the respective PWNe, and we
list some important parameters of both systems in Table~\ref{tab:g213c58}.

\begin{deluxetable}{lll}
\tablewidth{0pt}
\tablecaption{\label{tab:g213c58}Parameters of PWNe \snr, 3C~58, and
their Pulsars. }
\tablecolumns{3}
\tablehead{
\colhead{Parameter} &
\colhead{\snr}      &
\colhead{3C~58}
}
\startdata
Distance, $d$ (kpc)\dotfill                    & 4.7         & 3.2            \\
PWN size (pc$^2$)~\tablenotemark{a}\dotfill    & $1.8\times1.8$ & $10\times5$ \\
Termination shock radius, $r_s$ (pc)\dotfill   & $\la 0.05$  & $\approx 0.15$ \\
Radio luminosity ($10^7$--$10^{11}$\,Hz), ($10^{34}$\ergs)\dotfill & 1.4 &3.0 \\
Radio spectral index, $\alpha_r$\dotfill       &  0.0        & 0.1            \\
Break frequency, $\nu_b$ (GHz)\dotfill         & $\ga 500?$  & $\sim 50$      \\
Minimum internal energy, $E_{\rm min}$ (ergs)\dotfill & $\sim 2.5\times10^{47}$ & $\sim 10^{48}$ \\
X-ray luminosity (0.5--10\,keV), $L_X$ ($10^{35}$\,\ergs)\dotfill & 2.8 & 0.2 \\
X-ray spectral index, $\alpha_X$~\tablenotemark{b}\dotfill &0.4--1.3 &0.6--1.6\\
%\colrule
Equipartition magnetic field strength, $B_n$ (mG)\dotfill & $\approx 0.3$ & $\approx 0.08$ \\
Age, $\tau$ (kyr)\dotfill                      & $\la 1000?$ & 824?           \\
Pulsar period, $P$ (ms)\dotfill                & 61.8        & 65.6           \\
Spin-down luminosity, $\dot E$ ($10^{37}$\,\ergs)\dotfill & 3.3 & 2.7         \\
Characteristic age, $\tau_c$ (kyr)\dotfill     & 4.8         & 5.4            \\
Initial spin period, $P_0$ (ms)\dotfill        & $\ga 55?$   & $\sim 60?$     \\
\enddata
\tablecomments{We use the nominal distances listed here to infer all
quantities that depend thereon.  Parameters for \snr\ are from this
work and its references, while those for 3C~58 are from \citet{shvm04}
and references therein. }
\tablenotetext{a}{At least in projection, \snr\ is approximately round
(Fig.~\ref{fig:acis}) and 3C~58 is elongated. In both PWNe, the X-ray
nebula is somewhat smaller than the radio nebula, an effect more
pronounced in \snr\ \citep{bwd01}. }
\tablenotetext{b}{Both X-ray spectra steepen radially from the
center in similar fashion, within the ranges of $\alpha_X$ indicated
\citep{shp+01,shvm04,ms05}. }
\end{deluxetable}

Most noticeably, 3C~58 is a much larger and more asymmetric PWN than \snr,
and its ratio of X-ray luminosity to spin-down luminosity, $L_X/\dot{E}$,
is $\approx 11$ times smaller.  The inferred pulsar wind shock radii
differ by a factor of $\ga 3$, but this is understandable if the nebular
pressures differ by an order of magnitude, which may well be the case,
as inferred from the respective equipartition field estimates (see also
\S~\ref{sec:shock}).  Also, the break frequency $\nu_b$ for 3C~58 seems
to be $\sim 10$ times lower than for \snr\footnote{This may or may not be
of significance.  Spectral indices have not been reliably measured just
below and above the putative $\nu_b$ for either of these systems, and a
``break'' frequency is instead inferred from extrapolations of fluxes
measured 1--2 orders of magnitude apart.  The complexity of the broad
band emission from these PWNe may not be encapsulated by references to,
e.g., a single value of $\nu_b$.}.

3C~58 is widely thought to be the remnant of SN~1181~CE \citep{sg02b}.
In order to bring the historical age and measured spin parameters
into agreement, \citet{mss+02} proposed that the initial period of
the pulsar was $P_0 \sim 60$\,ms, comparable to what we infer for \psr\
(\S~\ref{sec:age}).  On the other hand, good arguments suggest that 3C~58
may be older \citep*{bkw01}, in which case $P_0 \ll 60$\,ms.  In any case,
different spin histories cannot explain different present-day values of
$L_X$, since the relevant electrons/positrons cool relatively quickly
in both PWNe.

If its apparently lower inferred $\nu_b$ and steeper X-ray spectrum are
of significance, much of the pulsar power in 3C~58 may be deposited
at lower particle energies, and hence radiated at lower frequencies,
compared to the case for \snr.  In addition, the lower nebular magnetic
field strength inferred for 3C~58 (see Table~\ref{tab:g213c58}) means
that its relativistic particles are relatively poor radiators, and this
may explain the vastly different X-ray efficiencies in the two systems.

But why is the nebular field strength in 3C~58 apparently so much
lower than for \snr?  If 3C~58 is not the remnant of SN~1181~CE but
is substantially older, as argued by \cite{bkw01} and \cite{che05},
then the larger PWN size would follow naturally, and its field would
also have decayed. Alternatively, even if the ages for both systems
are similar, the PWN radius at early stages depends on SN and pulsar
properties as $R \propto \dot{E}^{1/5} {E_{\rm SN}}^{3/10} {M_{\rm
ej}}^{-1/2}$, where $E_{\rm SN}$ and $M_{\rm ej}$ are the kinetic energy
and ejected mass, respectively, of the associated supernova explosion
\citep{rc84,vagt01}. Therefore, it may be possible to obtain a larger
size and lower magnetic field for 3C~58 if either the supernova explosion
was more energetic or the ejected mass smaller than the case for \snr,
either of which are reasonable.  Additionally, perhaps instabilities
in the magnetic field downstream of the termination shock are also of
relevance for explaining the elongated shape of 3C~58 and maybe its size
\citep[see, e.g.,][]{van03,shvm04}.

Finally, we note that the order of magnitude discrepancy in $L_X/\dot{E}$
between PSRs~J1833--1034 and J0205+6449, two pulsars with near-identical
spin parameters, makes it clear that the predictive power of simple
relations between $\dot E$ and $L_X$ for pulsars and their PWNe
\citep{sw88,bt97,pccm02} is limited.

\subsection{The Radio Spectrum of \psr\ and Young Pulsars}\label{sec:r-spectra}

The radio spectrum of \psr\ between 0.8\,GHz and 2.0\,GHz is unusually
steep ($\alpha \approx 3.0$; see \S~\ref{sec:r-conf}).  \citet{lylg95}
show that the mean spectral index for pulsars is $\approx 1.6$, with
a suggestion that it is smaller ($\sim 1$) for young pulsars.  The Crab
pulsar is an exception, with $\alpha = 3.1$.  \psr\ has a pulse duty cycle
of 4\%, smaller than virtually all pulsars with $P<0.1$\,s \citep[but
similar to that of J0205+6449 in 3C~58, which has similar period as
well, and $\alpha \sim 2$;][]{csl+02}.  With such a steep spectrum, the
pulsar is easier to detect at low frequencies, such as 820\,MHz at GBT.
However, for good reasons, the first frequency of choice to employ in
searching for distant pulsars along the Galactic plane remains 1400\,MHz.
At this frequency, \psr\ has a very low flux density and its luminosity
$L_{1400} \equiv S_{1400} d^2 \approx 1.8\,d_5^2$\,\mkpc\ is among the
smallest known.  The detection of such pulsars is therefore not easy.
Nevertheless, this may owe more to the state of our detection technology
than to the intrinsic rarity of such ``faint'' neutron stars.

We have compiled information about all known young rotation-powered
pulsars, defined here somewhat arbitrarily as those with $\tau_c <
30$\,kyr \cite[see the catalog of][but some of the distances used here
are from original references]{mhth05}.  There are 36 such pulsars, of
which 33 have been detected at radio wavelengths.  Of these, 23 have
a ``high luminosity'' $L_{1400}$ (18--140\,\mkpc; median 56\,\mkpc):
14 have been long known (of which three were first detected as X-ray
pulsars) and nine were discovered in the Parkes multibeam Galactic
plane survey \citep[e.g.,][]{kbm+03}.  With \psr, we know today of 10
young ``low-luminosity'' pulsars (0.5--6.3\,\mkpc; median 2.4\,\mkpc).
Seven of these were detected in deep searches of PWNe \citep[][and
this work]{hcg+01,mss+02,clb+02,cmgl02,cmg+02,csl+02,rhr+02}, two
more were detected in the Parkes multibeam survey (Camilo et al.\
2001, 2004)\nocite{cbm+01,cml+04}, and only one has been long known
\citep*[B1853+01, discovered in a directed search of SNR~W44;][]{wcd91}.
Evidently, such pulsars are not so rare, but merely require more effort
to uncover.

\section{SUMMARY}\label{sec:sum}

We have discovered the 61.8\,ms pulsar J1833--1034 at the center of the
composite SNR~\snr\ (\S~\ref{sec:r-pulse}).  Second only to the Crab in
spin-down luminosity among known rotation-powered Galactic pulsars, with
$\dot E = 3.3\times10^{37}$\,\ergs\ (\S~\ref{sec:r-conf}), \psr\ was born
in the SN explosion that gave rise to \snr, which we estimate from the
properties of the pulsar, PWN, and SNR shell occurred $\la 1000$ years
ago. This is substantially smaller than the pulsar characteristic age and
in the simplest interpretation implies a birth spin period of $\ga 55$\,ms
(\S~\ref{sec:age}), providing further evidence for a relatively large
range in initial periods for neutron stars \citep[e.g.,][]{mgb+02,klh+03}.

\psr\ is exceedingly weak at the radio frequency of 1.4\,GHz used at the
Parkes telescope to first detect it.  We have revisited the distance to
\snr\ based on \HI\ and CO observations, and find the best estimate to
be $4.7\pm0.4$\,kpc (\S~\ref{sec:dist}), for which the pulse-averaged
luminosity at 1.4\,GHz is about 1.5\,\mkpc, one of the smallest known
for a young pulsar. Observations at other frequencies with the GBT
(\S~\ref{sec:r-conf}) suggest that J1833--1034 has an unusually steep
spectrum, both of which findings have significant implications for
the detection of additional very young pulsars and for studies of the
radio luminosity, beaming fraction, and birth rate of such neutron stars
(\S~\ref{sec:r-spectra}).

We have analyzed a large number of \chandra\ X-ray observations of the
system and obtained the best image to date of the inner core of \snr.
This shows clear evidence for a pulsar component offset from the center
of a surrounding very compact elliptical emission region that likely
bounds the termination shock of the pulsar wind (\S~\ref{sec:x-image}).
From this, under the assumption of energy equipartition in the PWN, we
have obtained an estimate of the nebular magnetic field that is consistent
with other determinations and plausibly is also broadly compatible with
the X-ray and radio surface brightness and spectral energy distribution
observed in the PWN (\S~\ref{sec:shock}).  \psr\ has present-day spin
parameters very similar to those of PSR~J0205+6449 in SNR~3C~58, while
the respective PWNe exhibit some significantly different properties
(\S~\ref{sec:3c58}).

We have searched for X-ray pulsations from \psr\ without success
(\S~\ref{sec:x-pulse}).  Our pulsed fraction limit for an assumed
sinusoidal pulse shape is about 70\% of the point-like luminosity of
$3\times10^{-5}\dot E$.  Depending on the rotational stability of the
pulsar, to be determined from timing observations, we may in future
obtain a better limit or, if the X-ray beam intersects our line of sight,
a detection of such emission.  Interestingly, no EGRET $\gamma$-ray
source is known at the location of J1833--1034 \citep{hbb+99} ---
despite the fact that this neutron star has the fourth largest
spin-down flux ($\dot E/d^2$) among known rotation-powered pulsars
\citep[after the Crab, Vela, and J0205+6449, which also is not a known
$\gamma$-ray emitter;][]{csl+02}.  Assuming it beams toward the Earth,
its efficiency for converting rotational power into $>100$\,MeV $\gamma$
rays is relatively low, in keeping with a suggested inverse relation with
$\dot E$ \citep{tbb+99}.  Certainly, \psr\ should be a prime target for
the future {\em Gamma-ray Large Area Space Telescope}.

\acknowledgments

We thank Don Backer, David Kaplan, and Bryan Jacoby for their
contributions to developing pulsar observing equipment at GBT, Jereon
Stil for providing an absorption spectrum from the VLA Galactic Plane
Survey, and Tom Dame for providing CO data and useful advice.  We also
acknowledge useful discussions with Jules Halpern, Roger Chevalier and
Rino Bandiera.  We are grateful to the dedicated staff at Parkes and
GBT who make observing there such a productive pleasure.  The Parkes
Observatory is part of the Australia Telescope, which is funded by the
Commonwealth of Australia for operation as a National Facility managed
by CSIRO.  The National Radio Astronomy Observatory is a facility of
the National Science Foundation, operated under cooperative agreement
by Associated Universities, Inc.  We have made extensive use of NASA's
indispensable Astrophysics Data System.  FC acknowledges support from NSF,
NASA, and the NRAO travel fund.  BMG and POS acknowledge support from NASA
through grants NAG5-13032 and NAG5-9281, and through contract NAS8-39073.
DRL is a University Research Fellow funded by the Royal Society.

%\clearpage

\clearpage

\clearpage

\clearpage

\end{document}